# *GIER*: A Danish computer from 1961 with a role in the modern revolution of astronomy

*By Erik Høg, lektor emeritus, Niels Bohr Institute, Copenhagen*

**Abstract:** A Danish computer, *GIER,* from 1961 played a vital role in the development of a new method for astrometric measurement. This method, *photon counting astrometry,* ultimately led to two satellites with a significant role in the modern revolution of astronomy. A *GIER* was installed at the Hamburg Observatory in 1964 where it was used to implement the entirely new method for the measurement of stellar positions by means of a meridian circle, then the fundamental instrument of *astrometry*. An expedition to Perth in Western Australia with the instrument and the computer was a success. This method was also implemented in space in the first ever astrometric satellite *Hipparcos* launched by ESA in 1989. The *Hipparcos* results published in 1997 revolutionized astrometry with an impact in all branches of astronomy from the solar system and stellar structure to cosmic distances and the dynamics of the Milky Way. In turn, the results paved the way for a successor, the one million times more powerful *Gaia* astrometry satellite launched by ESA in 2013. Preparations for a Gaia successor in twenty years are making progress.

**Zusammenfassung:** Eine elektronische Rechenmaschine, *GIER*, von 1961 aus Dänischer Herkunft spielte eine vitale Rolle bei der Entwiklung einer neuen astrometrischen Messmethode. Diese Methode, *Astrometrie mit Photonenzählung* genannt, führte schliesslich zu zwei Satelliten, die grosse Bedeutung für die moderne Revolution der Astronomie bekamen. Eine *GIER* wurde 1964 in der Hamburger Sternwarte installiert, wo sie der Entwicklung der ganz neuen Methode zur Messung von Sternpositionen mithilfe eines Meridiankreises diente, damals das Hauptinstrument in der Wissenschaft der *Astrometrie*. Eine Expedition mit dem Instrument nach Perth in Westaustralien war ein Erfolg. Diese Methode wurde auch in dem ersten astrometrischen Satelliten, *Hipparcos*, verwendet, die von ESA im Jahre 1989 ins All geschossen wurde. Die Ergebnisse von *Hipparcos* wurden 1997 publiziert und in allen Gebieten der Astronomie verwendet, vom Sonnensystem und Struktur der Sterne bis zur kosmischen Entfernungsskala und Dynamik der Milchstrasse. Des Weiteren bahnten die Ergebnisse den Weg für einen Nachfolger, den *Gaia* Astrometriesatellit, der von ESA im Jahre 2013 gestartet wurde. Vorbereitungen für einen Gaia Nachfolger in zwanzig Jahren gehen voran.

## Introduction

After completing my studies at the University of Copenhagen where Bengt Strömgren (1908-1987) had been professor of astronomy, *Figure 2,* I arrived at the Hamburg Observatory in Bergedorf in October 1958 for my first professional stay at a foreign astronomical institute. The stay was financed by the *Deutsche Akademische Austauschdienst* and was planned to last for ten months. This turned out to be 15 years because my new ideas about astrometric measurements fitted so well with the plans for the Observatory that I was soon given a permanent position.

Through my stay in Hamburg I aimed to change my direction in astronomy from the field of astrometry to the field of *astrophysics*. Astronomy courses at Copenhagen University gave a broad background in most contemporary branches of astronomy but my main work had been focussed on astrometry, especially utilising the newly erected meridian circle, Figure 3. This was a special telescope designed to measure the position of stars as they cross the meridian one by one. *Astrometry* had been the main occupation of astronomers for centuries, in fact since Antiquity, but astrometry had gradually begun to look old fashioned when compared with astrophysics. Astrophysics was based on the theories of thermodynamics and radiation and atomic physics necessitating the development of ever larger telescopes over the past one hundred years.

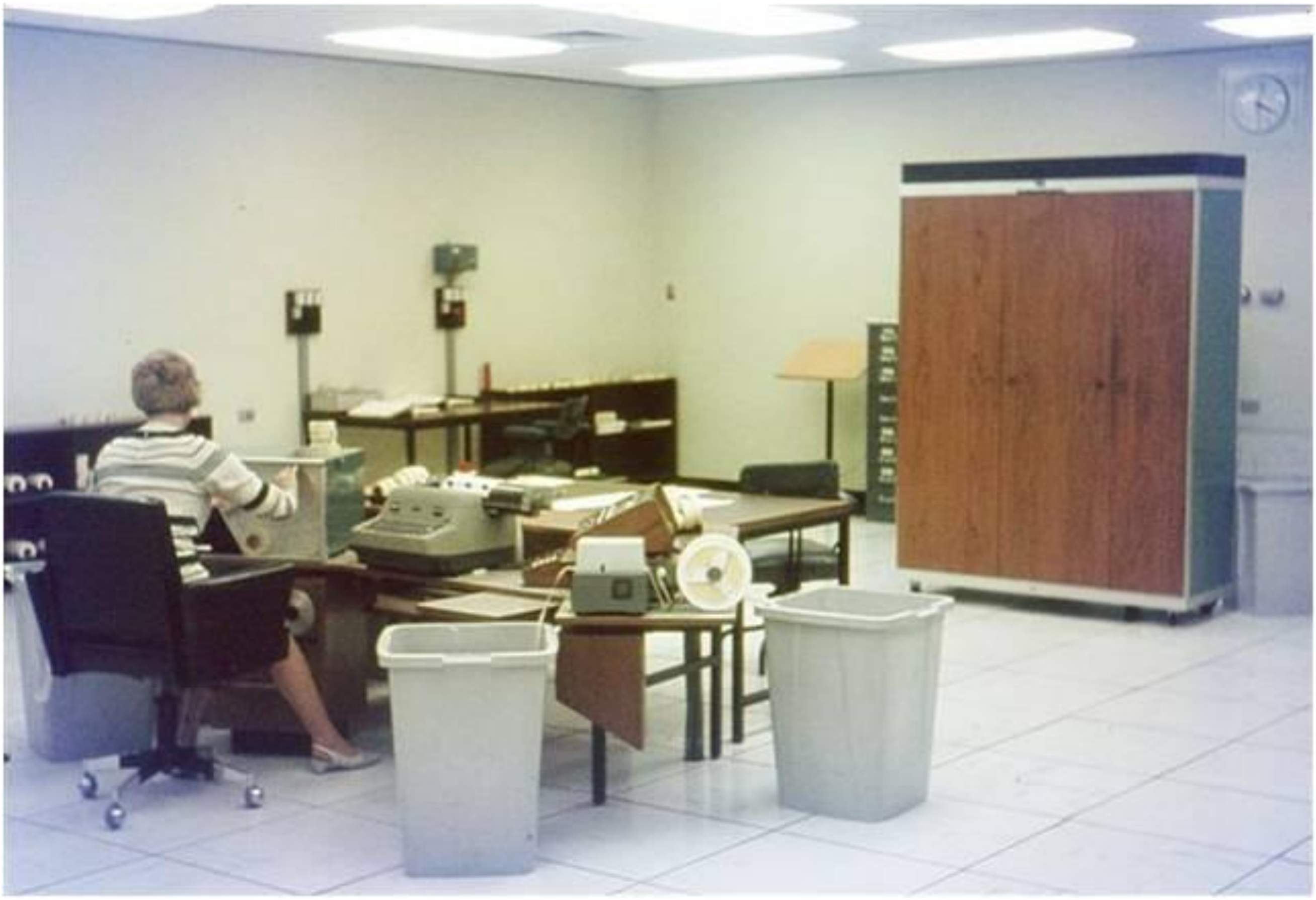

**Figure 1** *GIER* computer room of Perth Observatory 1971. *GIER*, in the teak cupboard in the background at right, was one of the first transistorized computers. Mrs Ilse Holst at the reader for 8-channel punched tape, one roll could contain 120 kilobytes. The control panel shown in *Figure 4* is seen just behind the paper punch. The tape from reader or punch poured into a large basket and could then be quickly rolled up. The operator would type commands to start a program on the typewriter and e.g. error messages from the computer were typed on endless paper. - Photo: Bernd Loibl.

My transition to astrophysics went reasonably well, but was interrupted in July 1960 when I had an idea for a new astrometric method described in *Figure 2*. The idea was so well received at the observatory and elsewhere that I returned to work in astrometry. Digital techniques and computers, especially the *GIER* computer, were crucial in the instrument development that followed. The method called *photon counting astrometry (Høg 1960)* was developed for the Hamburg meridian circle, a classical type of ground-based telescope for astrometry. The *GIER* computer was crucial for this development because it was just good enough to do the necessary calculations being ten times faster than any affordable alternative computer on the market. The development of astrometry at this time was crucially closely dependent on the development of computers and of computer science.

As explained below, the meridian circle and the computer were moved to Perth in Western Australia, *see Figure 1,* and later on I participated in the development of two astrometric satellites. This report is based on memory and on documents in *Hamburg Observatory (1969, 1970* and *1973*) as well as on previous papers, (e.g. *Høg 2014* and *2017*). Focus is on the computer issues, omitting a description of the instrument details easily available in *Høg (2014)* from where *Figures 1, 2,* and *3* are taken.

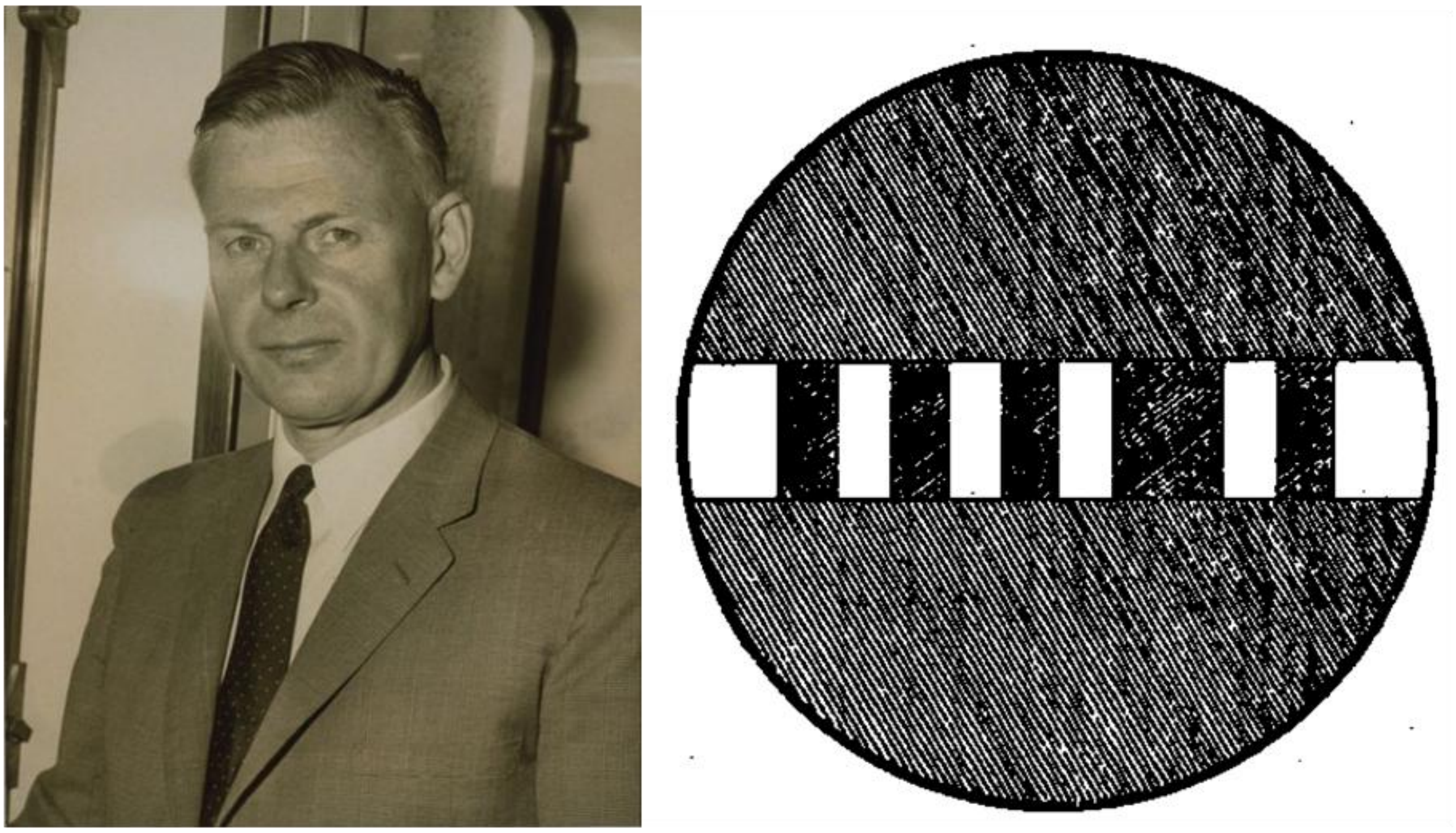

**Figure 2** Bengt Strömgren (1908-1987) in 1957, famous astrophysicist and professor of astronomy at Copenhagen University 1940-1951. On the right, the slit system he used starting in 1925, aged only 17, for his experiments in photoelectric recording of star transits (*Strömgren 1925*, *1926*, *1933*). A photocell behind the slits gave a signal in which the transit time for each slit could be detected thus obtaining the right ascension of the star. In 1960 my new idea was to replace the photocell with a photo multiplier tube and to use photon counting and digital recording on punched tape. I also introduced inclined slits so that two-dimensional measurement of the star was obtained, i.e. right ascension and declination *(Høg 1960)*. - Photos: Bengt Strömgren.

## Peter Naur and computers

Peter Naur (1928-2016) was my tutor at Copenhagen University from September 1953 and introduced me to astrometry, electronics and computing. Our computations in the 1950s were done manually with logarithms and mechanical desk calculators, but during his visits to Cambridge in 1951 and 1953 Naur had used the electronic computer *EDSAC*. After his visits to Cambridge, Naur gave a talk at Copenhagen University in a large auditorium which was so full that I had to sit on the floor. Niels Bohr (1885-1962) arrived late but was of course offered a chair. Naur had brought a roll of the then sensational punched tape which he threw to roll out into the astonished audience - and gave a brilliant talk.

My first job for Naur was to make interpolations in the tables he had brought back from *EDSAC*. Without Naur I would not so early in my career have become familiar with electronics and electronic computers which were outside the astronomical curriculum in most observatories. These were vital for my later work on astrometry. I had the good fortune to share office and workspace with Naur for several years, first in Copenhagen and then at the Brorfelde Observatory.

Naur laid the seeds of the great change of astrometry in two ways: firstly as my tutor and mentor and secondly, through his work on computer science which became crucial for me as I shall explain below. A European-American group developed the *ALGOL* programming language in those years and Peter Naur played a leading role *(Naur 1981)*. *ALGOL60* greatly facilitated my work on the computer programs for the Perth expedition since the high-level language was an immense step forward from the programming with numbers for the *IBM 650* I had used before. The *GIER ALGOL* compiler was called a masterpiece by Edsger W. Dijkstra (*Naur 1992* p.viii).

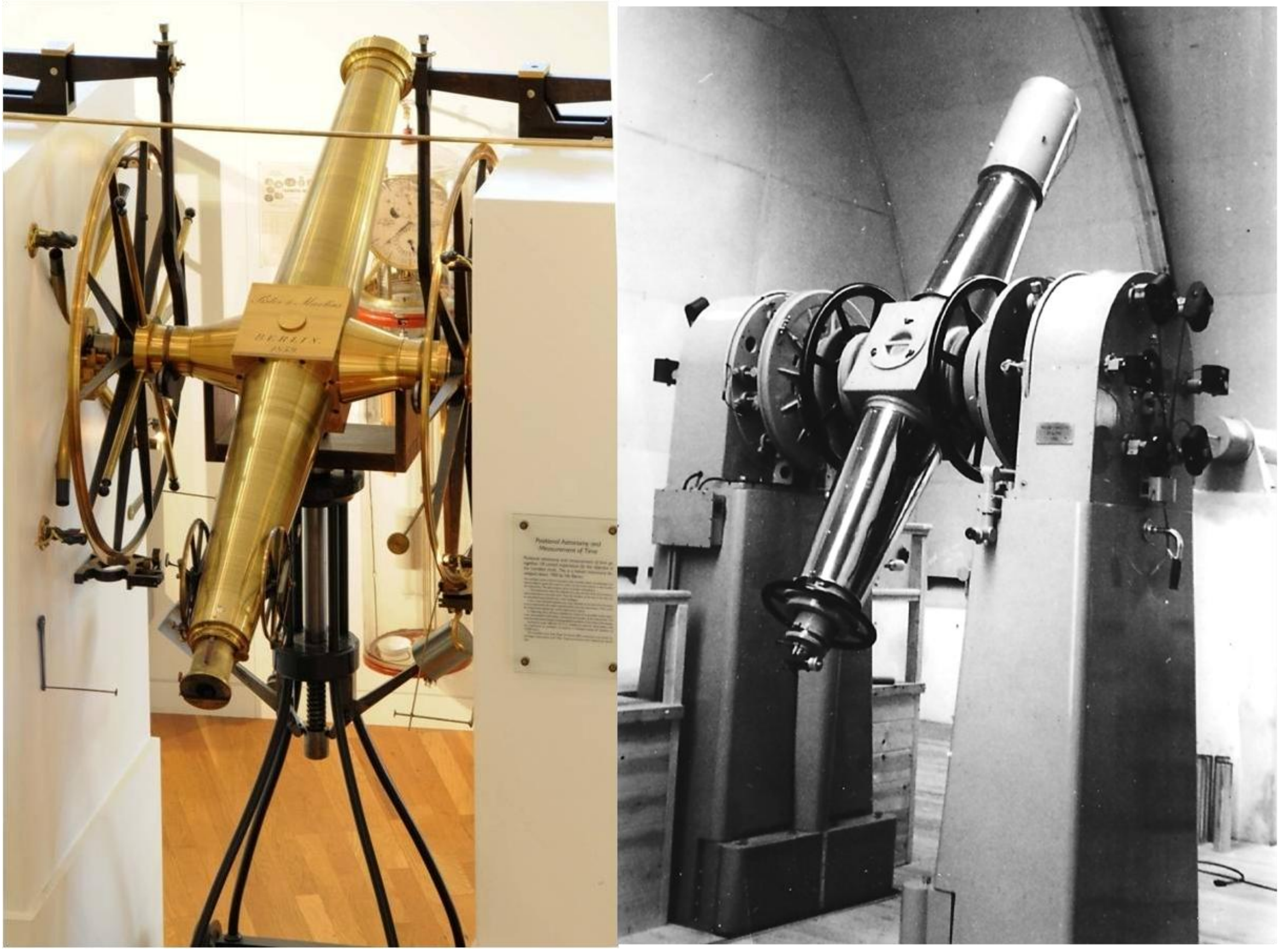

**Figure 3** Left: the Copenhagen meridian circle as used by Bengt Strömgren in 1925 – by Pistor & Martins, Berlin, 1859. Right: the new meridian circle by Grubb Parsons, Newcastle, installed in 1953 at Brorfelde, 50 km to the west of Copenhagen. The Brorfelde Observatory was established in 1953, away from the city light, dust and vibrations. – Photo on the left: Steno Museum, Aarhus, on the right: Erik Høg.

Bengt Strömgren was professor of astronomy at Copenhagen University until 1951 when he became director of the Yerkes and McDonald Observatories in the USA. In 1957 he became the first professor of astrophysics at the Institute for Advanced Study in Princeton, where he was given Albert Einstein's office. He did however still care deeply for astronomy in Copenhagen. He corresponded regularly and visited every summer. In December 1953, he engaged me to work with the new meridian circle in Brorfelde, and I reported regularly to him by letter about my observations. I admired him greatly for his contributions to science, for his perfect lectures and also for his personal appearance, always kind and willing to listen. It seemed to me in those years that a professor was someone like him – therefore I never had the ambition to become a professor, because I believed I could never be like him.

I was completely alone in Brorfelde at night and sometimes slept in a haystack when it became cloudy because there were no other buildings, only the meridian circle pavilion. A letter to Strömgren from the observatory director of that time, Julie Vinter Hansen (1890-1960), quotes Naur warning that such a task under such circumstances was likely to kill the interest in astronomy of any young man - I was 21 years old. But this did not happen, one reason being my already long dedication to astronomy and instrument making. (*See Høg 2017*).

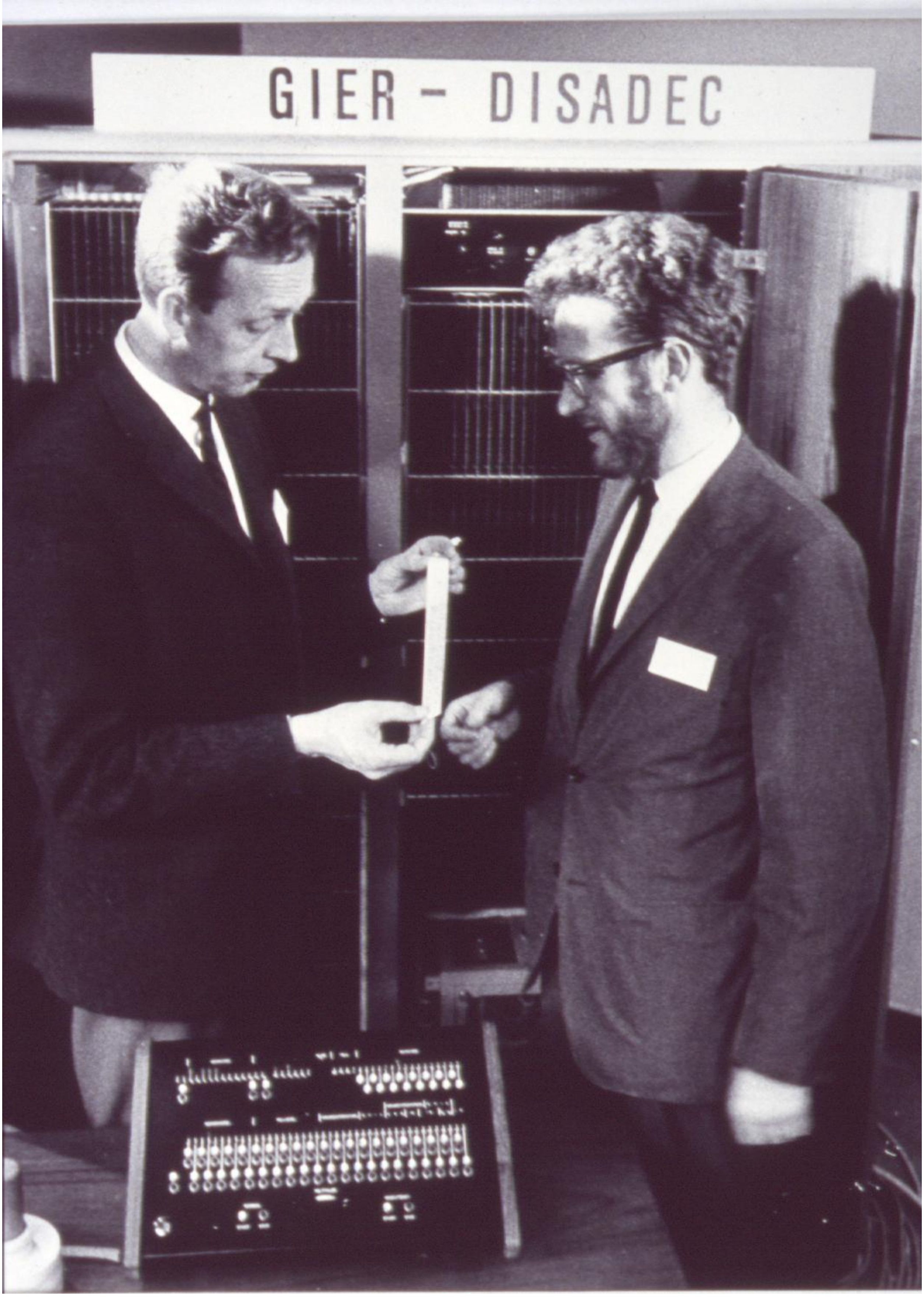

**Figure 4** Niels Ivar Bech (1920-1975), director of *Regnecentralen*, and Peter Naur in August 1962 in front of the *GIER* computer with the control panel on the table. - Photo: *Regnecentralen*.

In 1957 I became Naur's assistant at Brorfelde for one year. During that time I learnt to build electronics from scratch based on a circuit drawing, buying resistors, valves, capacitors etc. and soldering them for use at the new meridian circle. That knowledge soon became very useful in Hamburg-Bergedorf where it was something new. In 1958 Naur encouraged me to go abroad. I objected that I had to learn more before going, but he said that it was precisely now that I should go in order to learn more. My application to go abroad was approved and I left for the big observatory in Hamburg-Bergedorf in September 1958.

During the year at Brorfelde I enjoyed the hospitality of the Naur family at their home. Naur taught me to appreciate Opera, sitting beside me with the score, pointing at the notes while we listened to his LPs (long playing vinyl records) which were novel at the time and which he let me tape. He also brought me closer to George Bernard Shaw (1856-1950) the great Irish playwright, critic and political activist.

Peter Naur said in an interview with a Danish historian a few years ago that he was fed up with the Copenhagen Observatory. There was no leadership present and the work with the meridian circle for which he was in charge was merely instrument building and then trivial observation of stars, one by one. It was not science, only like land-surveying on the sky.

Naur abandoned astronomy in 1959 despite having been an astronomer since boyhood and completing his University degree at the age of 21. He joined *Regnecentralen*, a research institution under the Danish Academy of Technical Sciences. In 1969 he was called by the Copenhagen University to be the first full professor of computer science in Denmark (*Bruhn et al. 1988*). I visited him in 1978 and realized to my surprise that he had simply lost interest in astronomy. Late in life he again changed his field of interest turning away from computing, writing such works as *"How our mental life happens in our nervous system"* (*Naur 2015*). He won the 2005 Turing-award, the *"Nobel-prize of computing science"*. Naur died in 2016.

I enjoyed working with him. He did mention his frustrations to me but not so strongly in those years. It turned out, however, that I had been put on the right track towards instrumentation and astrometry, exactly right for my personality. I met him again in 2010 when he attended my lecture about the development of astrometry and I was happy when he, the teacher I had admired, later wrote to me: *"... unbelievable what you have achieved."*

Since Naur had meant so much to me I phoned him when I wrote about him for Section 10 in *Høg (2014)* and we had a very pleasant half-hour talk. He later read the draft and agreed. He told for instance that he had completely forgotten, actually repressed his three years in Brorfelde because they had been so unpleasant and scientifically unproductive for him – whilst Brorfelde had meant so much for me.

## Reception of the new idea for astrometry

The reception of my new idea for astrometry of 22 July 1960 is described in Section 4 of *Høg (2014)*. Briefly, it was soon clear that Otto Heckmann (1901-1983), professor and director of the observatory, wanted the idea implemented on the Repsold meridian circle. This instrument had been used to create the AGK2A catalogue, about 20,000 reference stars for the photographic AGK2 catalogue for which the photographic plates were taken in Hamburg and Bonn about 1932. The then young observer in 1932, Johann von der Heide (1902-95), was now of mature age and was in the years around 1960 observing the AGK3R, the reference stars for the photographic plates of AGK3 which were being measured in Bergedorf.

Von der Heide was far from being tired of all the individual meridian observations of thousands of stars. He was eager to lead an expedition with the meridian circle to Perth in Western Australia. The purpose was to observe the 20,000 Southern Reference Stars (SRS), an international project undertaken in 1958 with 13 participating telescopes in as many countries.

The intention was to observe in Perth with the Hamburg meridian circle precisely as it was equipped for the ongoing AGK3R program, i.e. with visual observation of the star. The director of Perth Observatory, H.S. Spigl, was erecting a new observatory at a good site far outside Perth in preparation of the coming Hamburg expedition.

Obviously, the expedition was a large undertaking for the Hamburg Observatory since six observers would be going to Perth to set up the instrument and run it for two years beginning in 1963. The decision to introduce a completely new and untried electronic technique instead of the classical visual observing was daring. It resulted in scientific success in the end, but put a great burden on the participants, especially because the work suffered delays: observations began in 1967 and lasted for five years instead of two and ten observers were involved during the last years instead of six as originally planned.

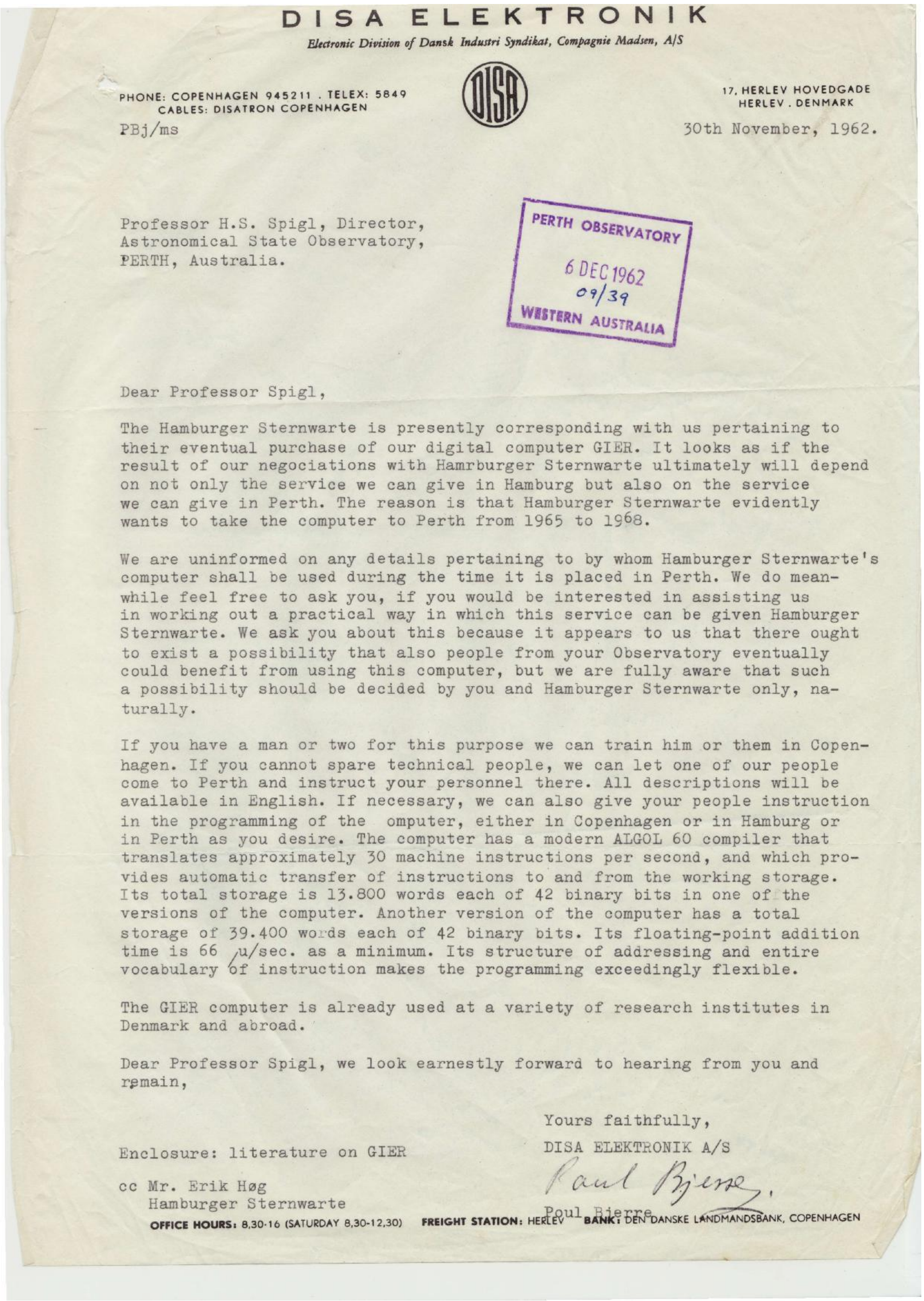

DISA ELEKTRONIK

Electronic Division of Dansk Industri Syndikat, Compagnie Madsen, A/S

PHONE: COPENHAGEN 945211 . TELEX: 5849
CABLES: DISATRON COPENHAGEN

DISA

17, HERLEV HOVEDGADE
HERLEV . DENMARK

PBj/ms

30th November, 1962.

Professor H.S. Spigl, Director,
Astronomical State Observatory,
PERTH, Australia.

PERTH OBSERVATORY
6 DEC 1962
09/39
WESTERN AUSTRALIA

Dear Professor Spigl,

The Hamburger Sternwarte is presently corresponding with us pertaining to their eventual purchase of our digital computer GIER. It looks as if the result of our negociations with Hamrburger Sternwarte ultimately will depend on not only the service we can give in Hamburg but also on the service we can give in Perth. The reason is that Hamburger Sternwarte evidently wants to take the computer to Perth from 1965 to 1968.

We are uninformed on any details pertaining to by whom Hamburger Sternwarte's computer shall be used during the time it is placed in Perth. We do meanwhile feel free to ask you, if you would be interested in assisting us in working out a practical way in which this service can be given Hamburger Sternwarte. We ask you about this because it appears to us that there ought to exist a possibility that also people from your Observatory eventually could benefit from using this computer, but we are fully aware that such a possibility should be decided by you and Hamburger Sternwarte only, naturally.

If you have a man or two for this purpose we can train him or them in Copenhagen. If you cannot spare technical people, we can let one of our people come to Perth and instruct your personnel there. All descriptions will be available in English. If necessary, we can also give your people instruction in the programming of the computer, either in Copenhagen or in Hamburg or in Perth as you desire. The computer has a modern ALGOL 60 compiler that translates approximately 30 machine instructions per second, and which provides automatic transfer of instructions to and from the working storage. Its total storage is 13.800 words each of 42 binary bits in one of the versions of the computer. Another version of the computer has a total storage of 39.400 words each of 42 binary bits. Its floating-point addition time is 66 μ/sec. as a minimum. Its structure of addressing and entire vocabulary of instruction makes the programming exceedingly flexible.

The GIER computer is already used at a variety of research institutes in Denmark and abroad.

Dear Professor Spigl, we look earnestly forward to hearing from you and remain,

Yours faithfully,
DISA ELEKTRONIK A/S
Paul Bjerre
Poul Bjerre

Enclosure: literature on GIER

cc Mr. Erik Høg
Hamburger Sternwarte

OFFICE HOURS: 8,30-16 (SATURDAY 8,30-12,30) FREIGHT STATION: HERLEV BANK: DEN DANSKE LANDMANDSBANK, COPENHAGEN

**Figure 5** A letter of November 1962 from *DISA Elektronik* representing *Regnecentralen* to the director of Perth Observatory, H.S. Spigl. Information on *GIER* is offered and the possible training of personnel from Perth Observatory on the use of *GIER* is discussed. - Photo: Craig Bowers.

## Danish computer beats American by a factor of ten

Now back to 1960 and the implementation of the proposed photoelectric system. The idea was published in *Høg (1960).* I defined the system for timing and recording on punched tape. On November 1, Heckmann sent an application to the Deutsche Forschungsgemeinschaft (DFG). Thus only three months after I had the first idea Heckmann had taken his decision and launched the project. This is documented in *Hamburg Observatory (1969)* and *(1970)* from which the following details are taken.

**The choice of computer**

There is no mention of a "computer" in Heckmann's application, it is only referred to once using the two words *"elektronische Rechenmaschine"*. With hindsight, this was very optimistic in 1960. Even three years later when the computer had to be chosen there was only one affordable computer on the market, which could do what we needed. It was *GIER*, Geodætisk Instituts Elektroniske Regnemaskine, produced by a small Danish firm, *Regnecentralen*, and it was ten times faster than its closest competitor produced by *IBM*. If we had been realistic in 1960 we might have concluded that the project was unfeasible, but our ignorance, enthusiasm and wishful thinking kept us going! – These sentiments are often useful drivers in new science.

I followed the markets for computers and electronics closely in those years. From visits to the annual industrial fair in Hanover, the biggest in the world (and only 150 km from Hamburg) I knew about electronic components and I had close contact to the companies. Agents from *IBM, Zuse, Standard Electric Lorenz* etc. came to see me in Bergedorf. But I did not, at this early stage, see how critical the issues of computer hardware and software and the development of the computer programmes were.

A plan in a letter from von der Heide to Heckmann of 18 April 1961 *(Høg 2012)* foresees observations during two years beginning in late 1963. It does not mention a "computer". The data from the meridian circle were punched on paper tape, which had to be analysed on a digital computer. In 1962 however when test observations were expected, the Hamburg University computer *TR4* would not be ready for another one or two years, especially with respect to reading paper tape. An agreement was therefore obtained with an institute in Bonn to use the *ER56* computer for free, and funds were given for 15 trips to Bonn. This solution would have required provisional programming in a special language different from *ALGOL*. Luckily this was never needed.

In September 1962 a *GIER* was installed at the Copenhagen Observatory, and I used it during a short visit on the 18th July 1963, assisted by the young student Richard West whom professor Anders Reiz had introduced me to. It became clear to me that this computer would be the best for our purpose. I also considered computers from *Zuse* and *IBM*, but none of them could match the *GIER* with respect to speed of computing and of handling punched tape or to the ease of programming with the *ALGOL* language, *see Figure 8* and *Naur (1963).*

A plan found in an application for funding of 28 August 1963 foresees again a 2-year observation period (identical to the 1961 forecast), only now it was to commence in 1965 at a total cost of 1.6 million DM, including salaries, a new pavilion in Perth and an *IBM 1620 model-2* computer.

In the end, the observations began in late 1967, took 5 years and a *GIER* computer was used. The *GIER* was 20 per cent cheaper and computed 20 times faster than the *IBM 1620.* The tape reading and punching were ten times faster. Using the *IBM 1620, see the Appendix, IBM (1959)* and *IBM (1962),* would have made the expedition unrealistic or impossible. I remember Gerhard Holst (1934-2000), first assistant to von der Heide, telling me that the observatory director Hans Haffner (1912-1977) wanted an *IBM 1620* and that I replied I would leave the project if that happened. Later on in Australia, it turned out that even with the *GIER* it was only merely possible to keep up with the analysis of all the observations.

I fully understand that Haffner and von der Heide favoured a computer from a big company like *IBM* which could presumably offer service as far away as Australia. Such service seemed impossible for a small Danish firm, but *Regnecentralen* did it with *GIER*! But how? The contract must have cost them a lot of money, placing a full time paid engineer in Australia during the whole period. Did they do it for the possible publicity? Or for the sake of astronomy?

We do not know what influenced their decision. In 2012 I asked the professors Peter Naur and Christian Gram who worked in or with the *Regnecentralen* in those years (though not in the marketing area) and they said that it was a general policy of the firm to target universities, thus reaching a whole new generation of users - students. Selling computers abroad and to renowned institutes meant a lot to the company. The letter in *Figure 5* gives an authentic glimpse of the selling techniques in 1962 of *Regnecentralen* represented by *DISA Elektronik*.

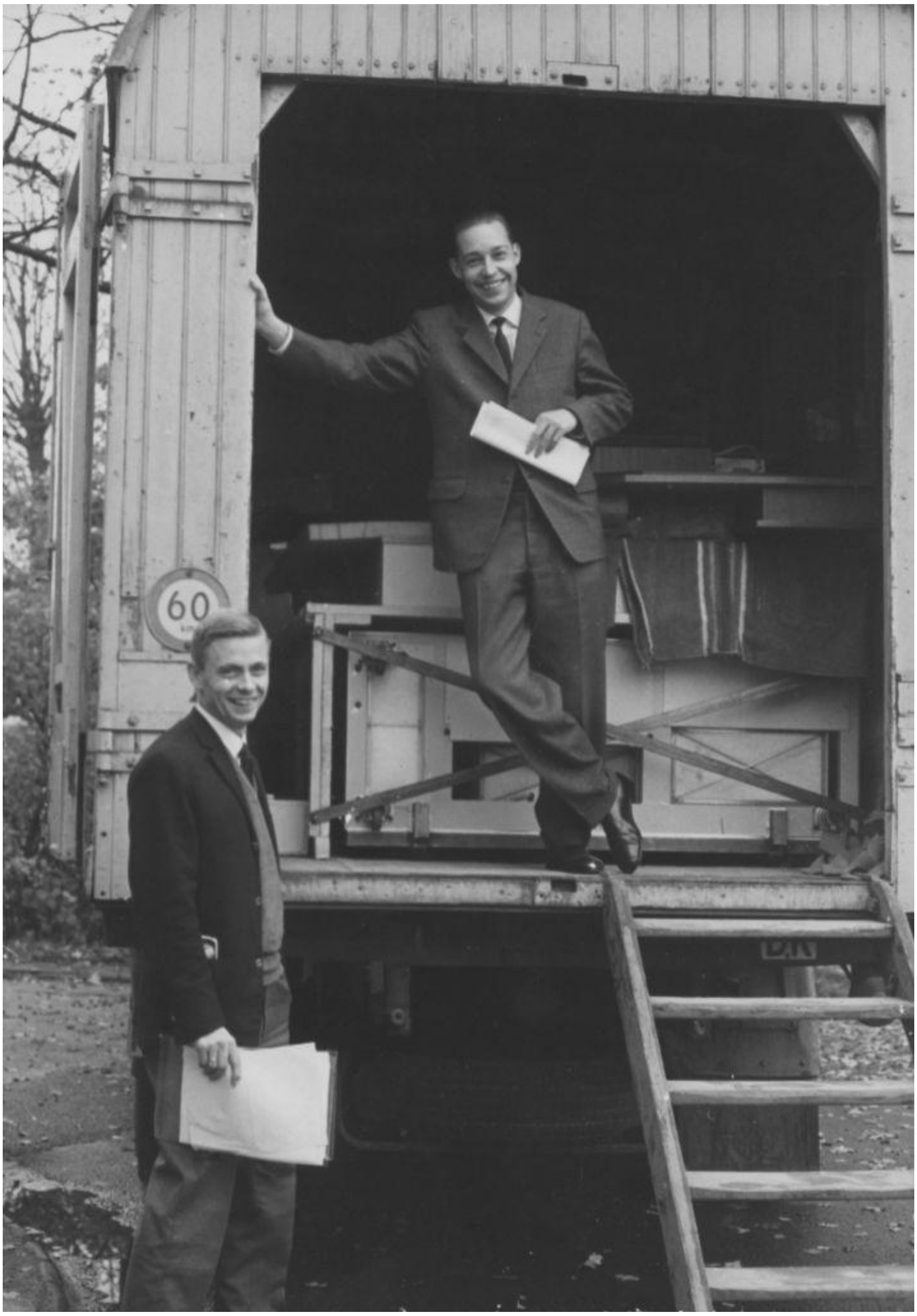

**Figure 6** *GIER* delivered to Hamburg Observatory in November 1964. At left Hans-Jørgen Hjort from *Regnecentralen*, in the lorry Gerhard Holst from the observatory. Mr Holst was a central person in the Australian expedition with great technical flair; without him it would not have succeeded. In Perth he also carried many administrative duties, hard working, very honest and modest as he was, cf. p.24 in *Høg (2014)*. - Photo: Wilhelm Dieckvoss.

**The *GIER* computer in Hamburg**

*GIER* is an acronym for *"Geodætisk Instituts Elektroniske Regnemaskine"* and it was taken into service at the Geodetic Institute on September 14th, 1961. The Copenhagen Observatory had a *GIER* installed in September 1962, initially with machine programming, but soon with an *ALGOL* compiler. In February 1964 this machine was replaced by one with improved design and mechanical construction.

The *GIER* computer, in the teak cupboard of *Figure 1,* was one of the first fully transistorized computers anywhere. It had the following approximate capacities: 5 kbytes RAM, 60 kbytes drum, and a computing speed of approx. 10 k-flops (flops = floating point operations per second). This was impressive at the time when we were used to computations by a desk calculator taking about 10 seconds for one multiplication of two number, i.e. 0.1 flops. But *GIER* had only a fraction of the capacity of any laptop computer of today. Interestingly, public authorities in Denmark with the Ministry of Finance in front thought in the 1950s that one electronic computer would be sufficient to serve all computing purposes of administrative nature in the whole country, a reasonable idea in those times. For research and education purposes more computers could be foreseen in those days.

The external storage of the Hamburg *GIER* was on 8-channel punched tape, one roll of tape could contain 120 kilobytes. Other versions of *GIER* were available with larger drums and disks and with magnetic tape, but the one on Hamburg was a minimal basic *GIER.* A tape could be read at the speed of 2000 characters per second which was unmatched in those years. Tape was punched at 150 characters per second.

The *GIER* was about ten times faster in tape input and output and in computing than the *IBM 1620* which had been considered in the above-mentioned letter of 28 August 1963. The *IBM 1620 model 2* could read 150 characters per second and punch 15 per second.

The first *GIER* in Germany was already installed at the Max-Planck Institute in Heidelberg. A small delegation from Hamburg Observatory visited and then travelled to Denmark, without my prior knowledge, to check the wisdom of my preference for the *GIER* computer. It was confirmed.

*GIER* was delivered in Bergedorf on a lorry in the afternoon of 2nd November 1964, *Figure 6*, and became operational the same evening. The price was 650,000 Danish kroner. We had the luck that a more powerful *ALGOL* compiler had become available shortly before our computer was delivered. *GIER* soon became the computer serving the whole observatory, easily accessible as it was in its location on the first floor, above the director's office. It was a much better option for all astronomers than to use the computer at the university 20 km away. It was used for 2.5 years, a total of 9300 hours for observatory purposes before being moved to Perth.

The annual report (*Hamburg Observatory 1973*) from Hamburg for 1964 states as follows:

*"Im Rahmen des SRS-Programms steht schliesslich die Beschaffung eines Elektronenrechners, der für die Dauer der Expedition nach Perth mitgenommen werden soll. In Zusammenarbeit mit der Deutschen Forschungsgemeinschaft wurde mit den Firmen Eurocomp, IBM, Zuse, Remington, CDC und Regnecentralen über eine für das SRS-Programm geeignete nicht zu teure Rechenanlage verhandelt. Die Entscheidung fiel auf die Rechenanlage GIER der Regnecentralen, nachdem Herr Holst im Mai bei der Gelegenheit einer Besprechung über einige Einzelfragen des SRS-Programms im ARI in Heidelberg eine beim dortigen Max-Planck Institut aufgestellte GIER-Anlage besichtigen konnte."*

***GIER* computer in Perth**

The Danish manufacturer, *Regnecentralen*, agreed to send an engineer, Jørgen Otto Sørensen, full time to Perth for the duration of the expedition to guarantee immediate repairs if needed. The firm must have chosen their best ever *GIER* because repair was only needed a few times during the five years. The engineer had nearly all his time for himself. According to Jørgen Otzen Petersen, who worked as astronomer at Copenhagen Observatory during those years, all *GIER* computers were very reliable. Sørensen stayed on in Australia with his family after the expedition. I used the *GIER* at my last visit in February 1980 and the last recorded use was on 22 September 1983.

Von der Heide had expected that the engineer would take part in the expedition also as observer, but that turned out to be a misunderstanding. I have not seen the contract where that should have been noted. Sørensen however took great care of the electronics for the meridian circle although he was sometimes hampered by severe health problems.

The astronomical results for 24,900 stars were published in a book by *Høg, von der Heide et al. (1976)*. Descriptions of the expedition and the instrumentation are contained in this book as respectively *Behr (1976)* and *Høg (1976)*. References to numerous papers on related issues are given in *Høg (2014)*. The result was one of the largest and most accurate contributions to the international SRS-Programme near completion in 1985 according to *Smith & Jackson (1985)*. There were 12 contributions obtained by visual observations and only the Perth observations were photoelectric.

But this respectable result was not to be rewarded by a continuation of modern meridian circle astronomy in Bergedorf. On the contrary, the observatory leadership, dominated by astrophysicists, decided to close the meridian circle for good, rejecting my proposal for an upgrade of the instrument to become fully automatic. Fortunately for me, I was offered tenure in Denmark and moved to Brorfelde in September 1973 where my experience with photoelectric photon counting astrometry was appreciated. Our method was implemented on the meridian circle there and I was soon called by ESA to join in the design of the first astrometric satellite *Hipparcos* which I did in 1975.

## The two astrometry satellites: *Hipparcos* and *Gaia*

From the beginning, French astronomers showed much interest in the photon counting astrometry proposed in *Høg (1960)* and in those days they called the slit system with inclined slits *"une grille de Høg"*, see *Figure 7* in *Høg (2014)* or p.23 in *Høg (2013)*. The annual meeting of French astronomers took place in Paris in June 1965 where I attended a *"Colloque Astronomie Fondamentale et Mecanique Celeste"* and gave two presentations: *"photoelectric measurement of star transits"* and *"photoelectric reading of declination circle"*. This was my first invitation to another country and all the other 17 presentations than mine in the two days meeting were in French as appears from the program just received from my colleague through all the years, Yves Réquième from Bordeaux Observatory.

Most importantly, however, the *photon counting method for astrometry* was crucial for the design of space astrometry, i.e. astrometry from a satellite, a visionary project being pursued by French astronomers since about 1964 and led by professor Pierre Lacroute (1906-1993), director of Strasbourg Observatory, see *Lacroute (1967, 1974)*. An ESA study group on this matter was established and held its first meeting in October 1975 in Paris. We were encouraged by the group leader to think independently from the very start on how we could use space technology for our scientific research. This enabled me to make a new design of an astrometric satellite in six weeks. The proposal was developed and the satellite was approved by ESA in 1980 as *Hipparcos, see Høg (2011a)*. It was launched in 1989 and observed for three years giving accurate astrometric and photometric results for 120,000 star in the *Hipparcos* Catalogue (*Perryman et al. 1997*) and 2.5 million stars in the *Tycho-2* Catalogue (*Høg et al. 2000*).

The *Hipparcos/Tycho* results caused a revolution in astrometry impacting on all branches of astronomy from studies of the solar system and stellar structure to the measurement of cosmic distances and the dynamics of the Milky Way. In turn, it paved the way for a successor, the one million times more powerful *Gaia* astrometry satellite launched by ESA in 2013.

The first results from *Gaia* for one billion stars were published in September 2016, confirming the high expectations, see *Gaia (2016)*. Publication of the first results only three years after launch, compared with the eight years it took for the *Hipparcos* results, has been possible thanks to the experience with *Hipparcos* and to the immensely powerful computers of today, thousands of times more powerful than those used for *Hipparcos* 25 years ago. One of the six processing centres for *Gaia* is the *ESAC* located near Madrid, it can do about 50 Tflops ~ 0.5 $10^{14}$ flops, according to Uwe Lammers in February 2017. The total processing effort during the entire mission to produce the final *Gaia* catalogue at the end of mission is estimated to be $10^{21}$ flop, an older estimate which still seems to hold.

On June 20, 2016, China's *Sunway TaihuLight* was ranked the world's fastest computer with 93 peta-flops ~ $10^{17}$ flops, according to Wikipedia *Sunway (2016)*, so that in principle it could process all *Gaia* mission observations in a few hours. *GIER* did 10 000 flops.

## Concluding remarks

It has been pleasant and thought-provoking for me to report on my early days during the 1950s and 60s in astrometry and computing and to draw the connection to my later work in astronomy in the very fruitful collaboration on *Hipparcos* and *Gaia*. All the time during the last 60 years new and fantastic possibilities have opened up for the development of astronomy and astrophysics, most recently the idea of a Gaia successor to be launched in twenty years (*Høg 2014b*), always driven by the wish to obtain higher astrometric accuracy and fainter stars and always in international co-operation. The results indicate that I have done more for astronomy by returning to astrometry in 1960 than I could have done as an astrophysicist.

ESA has in March 2017 approved our proposal (*Hobbs et al. 2016*) for a technological study of a detector with enhanced infrared sensitivity suited for a Gaia successor.

**Acknowledgements**: I am grateful to Finn Aaserud, Craig Bowers, Hans Buhl, Erik Frøkjær, Johan Fynbo, Christian Gram, Povl Høyer, Ole J. Knudsen, Uwe Lammers, Finn Verner Nielsen, Jørgen Otzen Petersen, Holger Pedersen, Yves Réquième, Caroline Soubiran, Rob Sunderland, and Gudrun Wolfschmidt for information and comments to previous versions of this report.

# Appendix: Small computers about 1960

## Regnecentralen

The following is based on *GIER* (1961) and *Gram (2011)* with a few corrections and additions, some due to Finn Verner Nielsen, employed by *Regnecentralen* from 1974 to 1983.

*Regnecentralen*, or *RC* for short, was the first Danish computer manufactorer, founded on October 12, 1955. They designed a series of computers, originally for their own use, and later to be sold commercially. Descendants of these systems sold well into the 1980s. They also developed a series of high-speed paper tape readers, and produced *Data General Nova* compatible machines.

**Genesis**

What would become *RC* started as an advisory board formed by the Danish *Akademiet for de Tekniske Videnskaber* (Academy of Applied Sciences) to keep abreast of developments in modern electronic computing devices taking place in other countries. After several years in the advisory role, in 1952 they decided to form a computing service bureau for Danish government, military and research uses. Led by Niels Ivar Bech (*Figure 4*) after 1957, the group was also given the details of the *BESK* machine being designed at the Swedish Mathematical Center (*Matematikmaskinnamndens Arbetsgrupp*).

The group decided to build their own version of the *BESK* to run the bureau, and formed *Regnecentralen* in October 1955 to complete and run the machine. The result was the *DASK*, a vacuum tube-based machine that was completed in 1957 and went into full operation in February 1958. *DASK* was followed in 1961 by the fully transistorized *GIER*, used for similar tasks. *GIER* is an acronym for *"Geodætisk Instituts Elektroniske Regnemaskine"* and was operational at the Geodetic Institute on September 14th, 1961.

Characteristic numbers are according to *Gram et al. (1963)*: 42-bit words, 1-k words core store, 12-k words drum store, 50 μs fixed point addition, 100 μs floating point addition, ie. 5 kbytes RAM, 60 kbytes drum and 10 k-flops (floating point operations per second).

# GIER SYSTEM

## SPEZIFIKATION

## ZENTRALEINHEIT

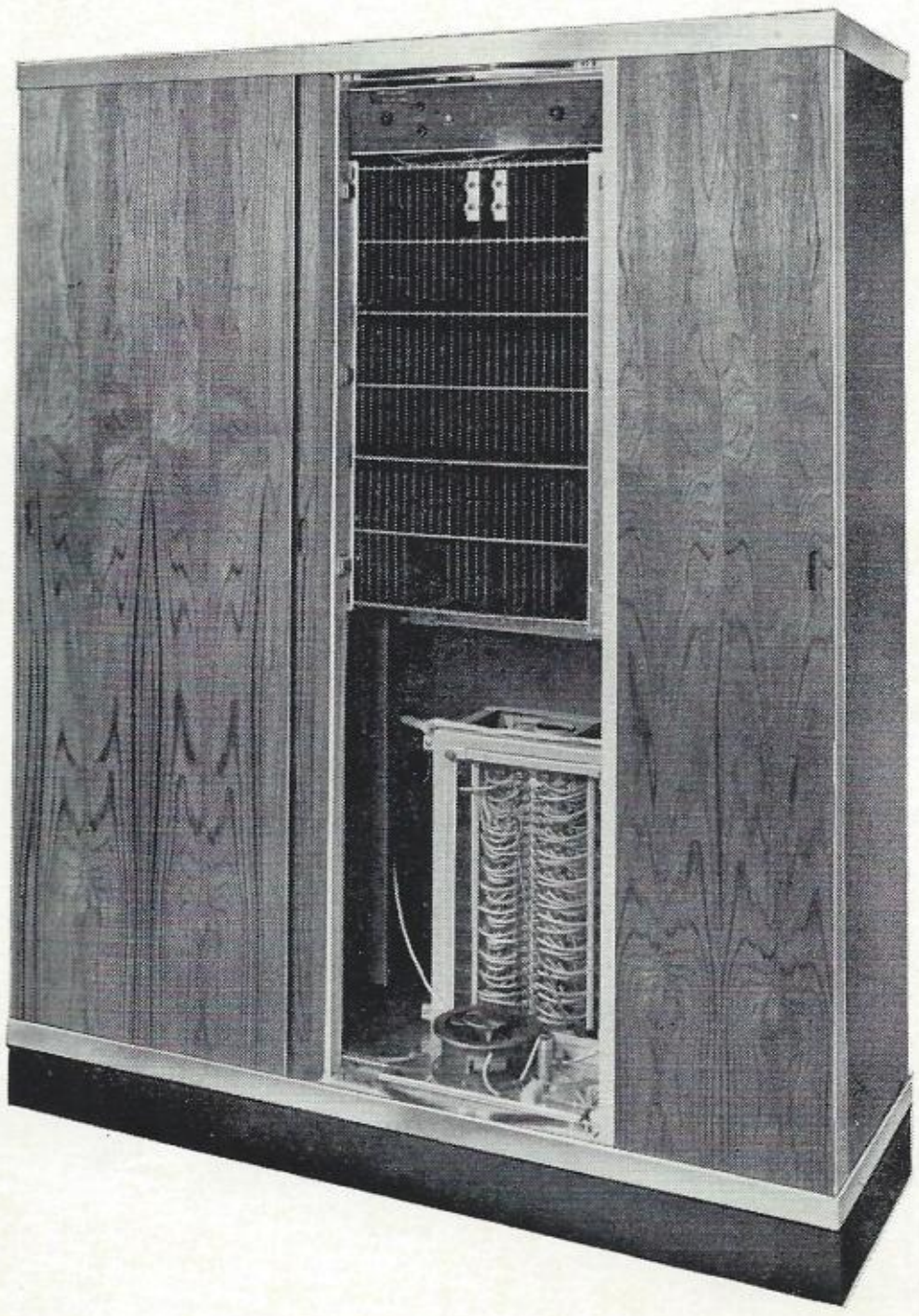

GIER ist eine binäre, parallele, Einadress-elektronische Rechenanlage mit eingebauter Gleitkomma-Arithmetik, automatischer Adressenmodifikation und Indexregister-Befehlen.

Die GIER-Zentraleinheit besteht aus einer Anzahl von Befehlregistern und einem schnellen Ferritkern-Speicher – dem IAS (immediate-access ferrite-core store) – mit 1024 Worten je von 42 Bits (jedes Wort entspricht 7 alphanumerischen Zeichen oder 12 Dezimalen). Die Zykluszeit ist 10 µs.

Zusätzlich zu IAS ist ein Trommelspeicher vorhanden, der 320 Spuren mit je 40 Worten hat (12.800 Worten), das heisst eine Gesamtkapazität von 13.824 Worten. Die Trommelübertragungen (Übertragung vom IAS zur Trommel oder umgekehrt) werden simultan mit den laufenden Rechnungen in GIER mit einer Geschwindigkeit von 21 ms. pro Spur vorgenommen. Durch den Anschluss von weiteren 2 Trommeln kann die Kapazität des Trommelspeichers auf ein Maximum von 38.400 Worten erweitert werden.

Um weitere externe Einheiten und direkte Nachrichtenverkehre zu ermöglichen, ist die Zentraleinheit mit einem Allzweck-Datenkanal und kann auch mit einer »Interrupt«-Einheit versehen werden. Die Information im Datenkanal kann zu und von dem IAS mit einer Geschwindigkeit von 5 µs pro Wort (42 Bits parallel) übertragen werden. Die »Interrupt Einheit« umfasst 12 »Interrupt-Kanäle« und ein dazu gehörendes »masking register«. »Interrupts« sind Signale, die die Rechenanlage auf interne oder externe Zustände aufmerksam machen, die entstanden sind und das augenblickliche Eingreifen der Rechenanlage erfordern. Ein »Interrupt«, der an einer »mask« vor- geht, wird verursachen, dass ein Programm, das ausgeführ »interrupted« (unterbrochen) wird. Ein spezielles Unterprogra kommt für eine gewisse Zeit den Vorrang, dann übernim Hauptprogramm wieder die Priorität.

Der IAS kann mit einem weiteren Speicher – dem Puffer – werden, auch ein Ferritkern-Speicher mit einer Kapazi Worten. Über den Datenkanal kann man an den IAS e von diesen Puffern mit einer Übertragungsgeschwin 6–13 µs./Wort anschliessen. Bis zu 4 Magnetbandeinhe über jedem Puffer an GIER angeschlossen werden.

| Operationszeit | Addition | Multiplikation |
|---|---|---|
| Festkomma | 49 µs | 180 µs |
| Gleitkomma | 93 µs | 170 µs |

| | Gewicht | Höhe | Breite | |
|---|---|---|---|---|
| Zentraleinheit | 500 kg | 193 cm | 144,6 cm | |
| Pufferspeicher | 400 kg | 193 cm | 144,6 cm | 54 |

1000-8-64

A/S REGNECENTRALEN VERKAUFSABTEILUNG: SMAL ... GADE 2 · KOPENHAGE... FERNSPRECHER: (01) FA 9916

**Figure 7** Specifications for *GIER* from 1964, in German; damaged at lower right. - From Regnecentralen.

1969 KRP 1

```
Kreisreduktion, KR herstellen
begin
comment prg: KRP 1, 2, MK 2, M10, M 11. daten: ausgang, binout, NL, kreisreg;
array M[1:6,1:6]; boolean array BNL,BKR[1:1];

procedure programproc;
begin integer i,j,m,anz; real v,a;
  writetext(≮< kreisreduktion , KR herstellen
 ≯);
  if kbon then gierproc(≮<binout≯,1);
  initialkr;
C:writecr;  writetext(≮<1  von=≯);
  von:=typein;  for i:=1,2,3,4 do M[6,i]:=0;   anz:=0;
  for iobs:=von step 1 until nobs do
    begin array q[1:na];
      snratio:=bowrite:=iobs-iobs:20×20=1;
      write(≮ndd≯,writecr,iobs);
      if skip>0 then skipuntil(11);
      drumplace:=plnl-iobs+1; fromdrum(BNL); split(BNL[1],0,14,nr);
goon: ch[1]:=0;  i:=0;
      for i:=i+1 while i<4 do
       begin   if M[1,5]=0 ∧i=2 then skipuntil(11);
comment vor 30.1.1967 (M[1,5]=0 ) wurde grobmikr. als 2. gemessen aber nicht
benutzt. Danach wird grob gar nicht gemessen;
          m:=i;    qplace:=drumplace:=M[3,m];
          nreg:=0;  readreg(abs(sig),q,al);
          if al then nreg:=0;  M[4,m]:=nreg;
          if last=10 then
           begin  readkenn(P,ch,al);  if al then ch[1]:=70000;
B:           if lyn=10 then
              begin writetext(≮<
neue kennz≯); if skip=0
               then
                begin readkenn(P,ch,al);if al then ch[1]:=70000;goto B end
               else
                begin readkenn(PB,chb,al);
                  ch[1]:= if al then 70000 else chb[2];  goto B;
                end if skip>0;
              end if neue kennziffern;
             for j:=2 step 1 until abs(sig) do  if lyn>9 then antwort(
```

**Figure 8** Part of an *ALGOL* program for the meridian circle in Perth. - From Erik Høg.

*GIER* proved to be a useful machine, and went on to be used at *RC*'s service centers, at universities, and several companies. Niels Ivar Bech also sold *GIER* machines to the Eastern Bloc nations, starting with Czechoslovakia, Hungary, Poland and Bulgaria. A daughter company *GIER* Electronics GmbH had head quarters in Hannover. *GIER* was also sold to Paris, Trondheim University, Heidelberg and to Hamburg from where it was moved and installed in Perth, Western Australia. About 45 machines were produced. *See Figure 7.*

*RC* was also home institute to Peter Naur, and *DASK* and *GIER* became particularly well known for their role in the development of the famous *ALGOL* programming language. After the first European *ALGOL* conference in 1959, *RC* started an effort to produce a series of compilers, completing one for the *DASK* in September 1961. A version for the *GIER* followed in August 1962. Christian Andersen, another *RC* employee, wrote the first introductory text on the language, Everyman's Desk *ALGOL*, in 1961. *See Figure 8.*

**Peripheral business**

In order to support higher productivity at their own service bureau, *RC* developed several high-speed input/output devices. One of their most popular was the *RC 2000* paper tape reader, introduced in 1963. Feeding the tape at 5 meters per second, the *2000* could read 2,000 characters per second (CPS), storing the results in a buffer while the computer periodically read the data back out instead of stopping the tape to wait for the computer to get ready. The machine was later upgraded as the *RC 2500*, increasing speed to almost 7 meters a second, equivalent to a read speed of 2500 CPS. The *RC 2000/2500* became a major product for *RC* during the 1960s, selling 1,500 units around the world. Several related devices were added to the line, including a high-speed *Facit* punch capable of punching 150 characters per second and a dedicated data conversion machine that would "massage" data between formats to ease the burden on the host computer.

In 1964 *Regnecentralen* was privatized, with the majority of the shares held by a broad range of large companies and organizations.

**New computers**

In the mid 1960s, *RC* started the design of a small integrated circuit (IC) based computer system for industrial control and automation needs, initially to fill a request by a Danish company to automate a chemical factory they were building in Poland. The *RC 4000* design emerged in 1966 and was completed for the factory the next year. When combined with appropriate peripherals, almost always including an *RC 2000* along with several re-branded devices from other companies, the *RC 4000* was a highly reliable minicomputer, and went on to be sold to institutes and two large power stations in Denmark and across Europe. The *RC 8000* from the mid 1970s used newer generation ICs to shrink the *RC 4000* into a single rack mount system. The last in the series, the *RC 9000*, further shrunk the machine and improved performance to about 4 MIPS, and was sold in versions that could run either *RC 8000* programs, or Unix.

The *RC 4000* is particularly famous for its operating system, developed by Per Brinch Hansen. Known as the *RC 4000* multiprogramming system, it is the first example of a system using an extremely simple kernel along with a variety of user-selected programs that built up the system as a whole. Today this concept is known as a microkernel, and efforts to correct for microkernels' poor performance formed the basis of most OS research through the 1970s and 1980s. Brinch Hansen also worked with Charles Simonyi and Peter Kraft on the *RC 4000*'s Real-time Control System.

*RC* also started selling the *Data General Nova* in 1970 as the *RC 7000*, later introducing their own updated version as the *RC 3600* the next year. This series filled a niche similar to the *RC 4000*, but for much smaller installations. The *RC 7000/RC 3600* became a fixture of many Danish schools and universities.

During the 1980s, *RC* produced the *RC 700 Piccolo* and *RC 759 Piccoline* systems, which were primarily sold to Danish schools (although some were sold to companies both in Denmark and abroad). The *Piccolo* was powered by the *Zilog Z80A* CPU, while the *Piccoline* was powered by the *Intel 80186* processor.

*Regnecentralen* was in 1989 acquired by *ICL, International Computers Limited*, where the name was changed to *RC International* until the end in 1992.

# IBM

*IBM (International Business Machines Corporation)* dominated the computer market for many decades selling thousands of machines for military, commercial and scientific customers. In 1953 *IBM* launched its *IBM 650* machine, see *IBM (1953)* based on punched card input/output. In 1959 *IBM* offered an "inexpensive" machine the *IBM 1620 model I*, followed by a *model II* in 1962 about which information is available in the following brief extracts from respectively *IBM (1959)* and *IBM (1962).*

**IBM 650**

The *IBM 650* Magnetic Drum Data-Processing Machine is one of *IBM*'s early computers, and the world’s first mass produced computer. It was announced in 1953 and in 1956 enhanced as the *IBM 650 RAMAC* with the addition of up to four disk storage units. Almost 2,000 systems were produced, the last in 1962. Support for the *650* and its component units was withdrawn in 1969.

The *IBM 650* was a two-address, bi-quinary coded decimal computer (both data and addresses were decimal), with memory on a rotating magnetic drum. Character support was provided by the input/output units converting alphabetical and special characters to/from a two-digit decimal code. The 650 was marketed to scientific and engineering users as well as to users of punched card machines who were upgrading from calculating punches, such as the *IBM 604*, to computers. Because of its relatively low cost and ease of programming, the *650* was used to pioneer a wide variety of applications, from modeling submarine crew performance to teaching high school and college students computer programming.

The *IBM 7070* (signed 10-digit decimal words), announced 1958, was expected to be a *"common successor to at least the 650 and the [IBM] 705"*. The *IBM 1620* (variable length decimal), introduced in 1959, addressed the lower end of the market. The *UNIVAC Solid State* (a two-address computer, signed 10-digit decimal words) was announced by *Sperry Rand* in December 1958 as a response to the *650*. None of these had a *650* compatible instruction set.

**IBM 1620 model I**

The *IBM 1620* was announced by *IBM* on October 21, 1959, and marketed as an inexpensive "scientific computer". After a total production of about two thousand machines, it was withdrawn on November 19, 1970. Modified versions of the *1620* were used as the CPU of the *IBM 1710* and *IBM 1720* Industrial Process Control Systems (making it the first digital computer considered reliable enough for realtime process control of factory equipment).

Being variable word length decimal, as opposed to fixed word length pure binary, made it an especially attractive first computer to learn on - and hundreds of thousands of students had their first experiences with a computer on the *IBM 1620*.

Core memory cycle times were 20 microseconds for the *Model I*, 10 microseconds for the *Model II* (about a thousand times slower than typical computer main memory in 2006).

**Architectural difficulties of model I**

Although the *IBM 1620*'s architecture was very popular in the scientific and engineering community, computer scientist Edsger Dijkstra pointed out several flaws in its design in *EWD37*, *"A review of the IBM 1620 data processing system"* (see reference #3 in *IBM 1959)*.

The successor to the *IBM 1620*, the *IBM 1130* introduced in 1965 was based on a totally different, 16 bit binary architecture. (The *1130* line retained one *1620* peripheral, the *1627* drum plotter.)

**IBM 1620 model II**

The *IBM 1620 Model II* (commonly called simply the *Model II*) was a vastly improved implementation, compared to the original *Model I,* of the *IBM 1620* scientific computer architecture. The cycle speed was raised to 100 kHz. 8 track paper tape was read at the rate of 150 characters a second (cps). The tape punch operated at a rate of 15 cps. Further details are given in *IBM (1962),* for instance:

It had basic ALU (arithmetic and logical) hardware for addition and subtraction, but multiplication was still done by table lookup in core memory. Multiplication used a 200 digit table. Rather than being an available option, as in the *Model I,* the divide hardware using a repeated subtraction algorithm, was built in. Floating point arithmetic was an available option, as were octal input/output, logical operations, and base conversion to/from decimal instructions.

The core memory that was freed by the replacement of the addition table with hardware was used for storage of two selectable "bands" of seven 5 digit index registers.

The console typewriter was replaced with a modified *Selectric* typewriter, which could type at 15.5 cps - a 55% improvement over the *Model I.*

The entire core memory was in the *IBM 1625* memory unit. Memory cycle time was halved compared to the *Model I*'s (internal or *1623* memory unit), to 10 µs (i.e., the cycle speed was raised to 100 kHz) by using faster cores. A *Memory Address Register Storage (MARS)* core memory read, clear, or write operation took 1.5 µs and each write operation was automatically (but not necessarily immediately) preceded by a read or clear operation of the same "register(s)" during the 10 µs memory cycle.

The processor clock speed was also doubled, to 2 MHz, which was still divided by 20 by a 10 position ring counter to provide the system timing/control signals.